\documentstyle[12pt]{article}
\def\ga{\gamma}
\def\de{\delta}

\def\th{\theta}

\def\rh{\rho}

\def\ph{\phi}

\def\fr#1#2{{{#1} \over {#2}}}
\def\prt{\partial}

\def\frac#1#2{{\textstyle{{#1}\over {#2}}}}

\def\lsim{\mathrel{\rlap{\lower4pt\hbox{\hskip1pt$\sim$}}
    \raise1pt\hbox{$<$}}}
\def\gsim{\mathrel{\rlap{\lower4pt\hbox{\hskip1pt$\sim$}}
    \raise1pt\hbox{$>$}}}
\def\sqr#1#2{{\vcenter{\vbox{\hrule height.#2pt
         \hbox{\vrule width.#2pt height#1pt \kern#1pt
         \vrule width.#2pt}
         \hrule height.#2pt}}}}

\newcommand{\beq}{\begin{equation}}
\newcommand{\eeq}{\end{equation}}
\newcommand{\bea}{\begin{eqnarray}}
\newcommand{\eea}{\end{eqnarray}}
\newcommand{\rf}[1]{(\ref{#1})}

\renewenvironment{thebibliography}[1]
 { \rm
   \begin{list}{\arabic{enumi}.}
    {\usecounter{enumi} \setlength{\parsep}{0pt}
     \setlength{\itemsep}{3pt} \settowidth{\labelwidth}{#1.}
     \sloppy
    }}{\end{list}}

\begin{document}
\titlepage

\begin{flushright}
{IUHET 245\\}
{LA-UR-93-1128\\}
%{hep-ph/9304250\\}
{February 1993\\}
\end{flushright}
\vglue 1cm

\begin{center}
{{\bf BOUNDING THE PHOTON MASS WITH SOLAR PROBES
%\footnote{\tenrm }
\\}
\vglue 1.0cm
{V. Alan Kosteleck\'y$^a$ and Michael Martin Nieto$^b$\\}
\bigskip
{\it $^a$Physics Department\\}
\medskip
{\it Indiana University\\}
\medskip
{\it Bloomington, IN 47405, U.S.A.\\}
\vglue 0.3cm
\bigskip
{\it $^b$Theory Division\\}
\medskip
{\it Los Alamos National Laboratory\\}
\medskip
{\it University of California\\}
\medskip
{\it Los Alamos, NM 87545, U.S.A.\\}

\vglue 0.8cm
%{\tenrm ABSTRACT}
}
\vglue 0.3cm

\end{center}

{\rightskip=3pc\leftskip=3pc\noindent%\tenrm
Proposed close-encounter solar probes (Vulcan and Solar Probe)
are planned to have highly eccentric orbits,
with a perihelion of about $4R_S$
and an inclination close to $90^\circ$
out of the plane of the ecliptic.
We show this could allow at least an order-of-magnitude
improvement in the present directly measured limit on the photon mass.

}

\vfill
\newpage

\baselineskip=20pt

If the photon has a mass $m_\ga$,
the Maxwell equations are replaced with the Proca equations
\cite{pr}.
In Gaussian units, these are
\beq
\prt_\mu F^{\mu\nu} + \mu^2 A^{\nu} = \fr{4\pi}{c}J^{\nu}
\quad ,
\label{a}
\eeq
where
$F^{\mu\nu} = \prt^\mu A^\nu - \prt^\nu A^\mu$,
the parameter $\mu = m_\ga c/\hbar$ is the inverse Compton wavelength
of the photon,
and the four-vector potential $A^\mu= (\ph, \bf A)$
is comprised, as usual,
of the electric scalar potential $\ph$
and the magnetic vector potential $\bf A$.
For detailed discussions of these equations and their implications,
see Refs.
\cite{gn1}.
Current conservation, $\prt_\mu J^\mu=0$,
enforces the Lorentz condition $\prt_\mu A^\mu = 0$
and implies the breaking of gauge invariance,
which under certain circumstances
might be realized by spontaneous symmetry breaking
(see Ref. \cite{ssb} and references therein).
Alternatively,
a small photon mass may be generated in the context of string theory
\cite{ks}.
This paper describes a possible order-of-magnitude improvement
in the present directly measured bound on the photon mass.
For indirect limits, the reader is referred to the
discussion and citations contained in \cite{gn1}.

The best laboratory limit on $\mu$ comes from extending
the concentric-sphere test
of Coulomb's law.
The basic idea involves two concentric conducting spherical shells of
radii $a$ and $b<a$ with a potential $V$ applied to the outer shell.
In general, in the presence of a nonzero photon mass,
the electrostatic potential $\ph$ becomes a solution
to the modified Poisson equation
\beq
(\nabla^2 - \mu^2)\ph(\bf x) = -4\pi \rh(\bf x)
\quad ,
\label{b}
\eeq
subject to appropriate boundary conditions.
Writing $r = |\bf x |$, the solution between the concentric spheres is
\beq
\ph({\bf x}) = V \fr a r \fr {\sinh \mu r} {\sinh \mu a}
\quad .
\label{c}
\eeq
The potential difference between the two spheres is no longer zero.
For $\mu a \ll 1$,
\beq
\fr{\phi(a) - \phi(b)}{\phi(a)} \simeq
\fr{1}{6} \mu^2(a^2-b^2) + {\cal O}[(\mu a)^4]
\quad .
\label{d}
\eeq
Using a multi-shell application of this method,
Ref. \cite{wfh} obtained the bound
\beq
\mu \leq 6 \times 10^{-10} ~~{\rm cm}^{-1}
\equiv 10^{-14} ~~{\rm eV} \equiv 2 \times 10^{-47} ~~{\rm g}
\quad .
\label{e}
\eeq

Note that in Eq.\ \rf{d} the lowest-order
physical effects of a nonzero photon mass appear at order
$(\mu L)^2$, where $L$ is a length scale.
This result is general
\cite{gn1},
and it implies that an improved bound
can be obtained either by a more precise measurement
or by using a larger system.
In particular,
it suggests a consideration of the effects of a nonzero photon mass
on magnetic field lines.

The field from a point magnetic dipole ${\bf D} = D {\bf{\hat z}}$
at the origin is,
in polar coordinates with $\bf{\hat r} \cdot {\bf \hat z} = \cos \th$,
\beq
{\bf B}(r, \th) = D \fr{e^{-\mu r}}{r^3}
\left[\left( 1 + \mu r+\frac{1}{3}\mu^2r^2\right)
     (3{\bf \hat r} \cos \th -{\bf \hat z})
   -\frac{2}{3}\mu^2r^2{\bf \hat z}\right]
\quad .
\label{f}
\eeq
This shows that the presence of a nonzero photon mass
rescales the usual dipole field and introduces an additional term.
On a sphere surrounding the dipole,
the latter appears as a constant field antiparallel to the dipole.

In 1943, Schr\"odinger suggested
\cite{s}
using the earth's dipole field to limit $\mu$.
A modern analysis using satellite and ground-based observations
provides a conservative bound of
\cite{gn2}
\beq
\mu \leq 10^{-10} ~~{\rm cm}^{-1}
\quad .
\label{g}
\eeq
This is an improvement over the best laboratory bound because
the increased size of the earth dominates over the reduced precision
obtainable in measuring the field.

More generally,
suppose a magnetic dipole is found to have strength $S$
on  the equator of a sphere
of radius $R$ centered about the dipole.
If the smallest measurable field is $\de$ and if no constant
antiparallel field is observed,
then Eq.\ \rf{f} provides a bound on the photon mass of
\beq
\mu \lsim \sqrt{\fr{3\de}{2S}} \fr 1 R
\quad .
\label{h}
\eeq
This shows that an increase of a factor of 10 in the length scale
of an experiment is equivalent to an improvement
of a factor of 100 in the
sensitivity.
The best bounds on the photon mass can therefore be obtained by going
to larger systems.

Consider,
for example,
the planet Jupiter
with radius $R_J \simeq 7 \times 10^4$ km.
Typical magnetometers aboard interplanetary probes have sensitivities
$\de /S \simeq 10^{-4}$,
so repeated measurements along a trajectory at about 10$R_J$ could in
principle yield a bound of $\mu \lsim 10^{-13}$ cm$^{-1}$.
In practice,
there are substantial complications from the nature of the jovian
magnetosphere
(see, for example, \cite{d,g}).
Only within the inner magnetosphere (up to about $10R_J$)
does the planetary field dominate over effects from
external currents and the solar wind.
Even within this region,
there are contributions from external multipoles.
Moreover,
the planetary field itself has significant
quadrupole and octupole moments
and is tilted with respect to the axis of rotation.
These and other complications
prevent the attainment of the ideal bound suggested by
Eq.\ \rf{h}.
Using data from the Pioneer 10 flyby of Jupiter
and fitting higher multipoles to the field in the inner magnetosphere,
Ref. \cite{dgn} derived a conservative bound of
\beq
\mu \leq 2 \times 10^{-11} ~~{\rm cm}^{-1}
\quad .
\label{i}
\eeq

In addition to Pioneer 10,
there were three other Jupiter flyby missions in the 1970s:
Pioneer 11, and Voyagers 1 and 2.
Of these,
the best trajectory for our purposes was that of Pioneer 11,
which attained relatively high jovigraphic latitudes and a perijove
of $1.6R_J$.
The point is that to extract a good bound from Eq.\ \rf{f}
a clean separation of the usual dipole-type field
from the additional term is useful,
and this requires information about the field at latitudes
away from the equator.
Multipole fits to the planetary field using the newer data
obtained could generate an improvement in the bound \rf{i} by
a factor of perhaps two.

A further small improvement
might eventually be made by incorporating multipole fits
to data from the 1992 Ulysses encounter with Jupiter,
for which the outbound pass attained relatively high southern latitudes
\cite{u1}.
Another possibility in the future is the Galileo probe
\cite{gal},
due to arrive at Jupiter in 1995.
The orbiter portion of this craft will make several passes
at distances ranging from an initial perijove of
about $4 R_J$ to varying apojoves of about $(100 \pm 25) R_J$,
before entering the `tail-petal' orbit with apojove of
about $150 R_J$.
However,
the alignment of the orbits and their extension
beyond the inner magnetosphere make unlikely a further improvement of
much more than a factor of two in the bound on $\mu$.

The Sun provides another interesting opportunity
to decrease further the upper limit on $\mu$.
Its radius is $R_S \simeq 7\times 10^5$ km $\simeq 10 R_J$,
which means any experiments would involve much larger length
scales and so would access smaller values of $\mu$.
However, the solar magnetic field is more complex and less
well understood than that of Jupiter
(see, for example, \cite{m}).
There are additional complications stemming from such features
as sunspots and coronal holes.
Even the basic effects of solar rotation
such as the Archimedes spiral
\cite{p}
remain a subject of active research
\cite{jk}.
Moreover,
the dipole field itself is time-dependent,
changing strength, tilt, and even orientation
with the solar cycle.

Energetics have so far limited solar
probe trajectories to within a few degrees latitude of the ecliptic,
for which the dipole field is likely to dominate only
well within the closest perihelia of about $65R_S$
achieved by the Helios 1 and 2 missions.
An improved bound on $\mu$ could, however,
be obtained by a solar polar mission.
A probe with a trajectory inclined near $90^\circ$ to the
ecliptic offers the advantage of better
separations between the dipole and the added terms
in Eq.\ \rf{f} and between analogous terms for
higher-order multipoles.
It also implies passage over the polar region,
where it is likely that the magnetosphere is simpler
because the solar wind and the rotation rate are reduced.

The solar dipole moment at solar minimum is believed to be
about (3 G) $R_S^3$, to within a factor of three.
Given that current weak field magnetometers are sensitive
to fields of order $10^{-7}$ G,
a probe would need to have a perihelion of no less than
one or two A.U. if it is even to detect the solar dipole moment.
One solar polar mission, Ulysses, is currently underway
\cite{u2}.
In 1994, the Ulysses explorer will reach the south polar region
of the Sun in an orbit at an inclination of about $80^\circ$
to the ecliptic.
It will spend about eight months above $70^\circ$
heliographic latitude
and will pass over the poles at distances of about 2 A.U.
However,
given the complications likely to be present in the magnetosphere,
even over the poles,
placing a bound on $\mu$
probably requires a perihelion an order of magnitude smaller than this,
within about $20 R_S$.

During the first decade of the next century,
close-approach solar probes are envisioned both
by NASA (Solar Probe)
\cite{sp}
and by ESA (Vulcan)
\cite{v1,v2}.
These missions are planned to have highly eccentric orbits,
with a perihelion limited
to about $4 R_S$
by modern heat-shielding technology.
To maximize heliographic latitude coverage,
the orbits should be inclined at about $90^\circ$
to the ecliptic.
Antenna-pointing requirements suggest a period commensurate
with half an earth year.

As an example,
consider an operational orbit with a period of one earth year,
an aphelion of $1.981$ A.U.,
and an eccentricity 0.981
\cite{v3}.
Then, the distances from the probe
to the center of the Sun would be approximately $6 R_S$,
$8 R_S$, and $12 R_S$,
at $20^\circ$ before, directly over,
and $20^\circ$ past closest-polar approach, respectively.
With a trajectory of this sort,
the anticipated quiet conditions near solar minimum
and the broad latitude coverage of the orbit
should obviate difficulties
arising from the local swamping of the solar dipole field
by the fields due to coronal holes and sunspots.
With a precision $\de /S \simeq 10^{-4}$
and data at a distance of about $10 R_S$,
Eq.\ \rf{h} suggests that a bound as low as
$\mu \leq 10^{-14}$ cm$^{-1}$
might in principle be obtained.
It is certainly plausible that repeated orbits
could allow an improved photon-mass bound of
$\mu \leq 10^{-12}$ cm$^{-1}$.

We thank many colleagues for their helpful comments and observations.
They include:
A.N. Cox,  W.C. Feldman, H. Hill,  J.G. Hills, T. Hoeksema,
H.R. Johnson, J.R. Jokipii, E.J. Smith, A. Spallicci, and G.J. Tuck.
This work is supported in part by the United States Department
of Energy.

\end{document}